\documentclass{aa}  
\usepackage{graphicx}
\usepackage{txfonts}
\usepackage{natbib}
\begin{document}
\title{Kinetic simulation of the electron-cyclotron maser instability:\\ 
effect of a finite source size}
\author{A.A. Kuznetsov\inst{\ref{armagh},\ref{iszf}} \and
        V.G. Vlasov\inst{\ref{istu}}}
\titlerunning{Simulation of the electron-cyclotron maser instability}
\authorrunning{A.A. Kuznetsov and V.G. Vlasov}        
\institute{Armagh Observatory, Armagh BT61 9DG, Northern Ireland\\
           \email{aku@arm.ac.uk}\label{armagh} \and
           Institute of Solar-Terrestrial Physics, Irkutsk 664033, Russia\label{iszf} \and
           Irkutsk State Technical University, Irkutsk 664074, Russia\\
           \email{vlasov@istu.edu}\label{istu}}
\date{Received *; accepted *}
\abstract%
{The electron-cyclotron maser instability is widespread in the Universe, producing, e.g., radio emission of the magnetized planets and cool substellar objects. Diagnosing the parameters of astrophysical radio sources requires comprehensive nonlinear simulations of the radiation process taking into account the source geometry.}%
{We simulate the electron-cyclotron maser instability (i.e., the amplification of electromagnetic waves and the relaxation of an unstable electron distribution) in a very low-beta plasma. The model used takes into account the radiation escape from the source region and the particle flow through this region.}%
{We developed a kinetic code to simulate the time evolution of an electron distribution in a radio emission source. The model includes the terms describing the particle injection to and escape from the emission source region. The spatial escape of the emission from the source is taken into account by using a finite amplification time. The unstable electron distribution of the horseshoe type is considered. A number of simulations were performed for different parameter sets typical of the magnetospheres of planets and ultracool dwarfs.}%
{The generated emission (corresponding to the fundamental extraordinary mode) has a frequency close to the electron cyclotron frequency and propagates across the magnetic field. Shortly after the onset of a simulation, the electron distribution reaches a quasi-stationary state. If the emission source region is relatively small, the resulting electron distribution is similar to that of the injected electrons and the emission intensity is low. In larger sources, the electron distribution may become nearly flat due to the wave-particle interaction, while the conversion efficiency of the particle energy flux into waves reaches $10-20\%$. We found good agreement of our model with the in situ observations in the source regions of auroral radio emissions of the Earth and Saturn. The expected characteristics of the electron distributions in the magnetospheres of ultracool dwarfs were obtained.}%
{} 
\keywords{radiation mechanisms: non-thermal -- planets and satellites: aurorae -- stars: brown dwarfs -- stars: low-mass -- radio continuum: stars -- radio continuum: planetary systems}
\maketitle

\section{Introduction}\label{introduction}
The electron-cyclotron maser instability (ECMI) is believed to be responsible for generation of auroral radio emissions from the magnetized planets of the Solar System \citep{wu79, zar98, tre06}. It is very likely that similar radio emissions are also typical of exoplanets \citep{bas00, zar01, zar07, hes11}, although the sensitivity of the existing instruments does not allow us to detect such radio signals. It was recently discovered that a number of very low-mass stars and brown dwarfs (collectively known as ultracool dwarfs) are the sources of intense periodic radio bursts in the GHz frequency range \citep{hal06, hal07, hal08, ber08, ber09}; these bursts seem to be produced due to the ECMI in a way similar to the planetary auroral radio emissions, but in a much stronger magnetic field \citep{kuz11b}. The electron-cyclotron maser has been also suggested as the generation mechanism of the periodic radio bursts from the magnetic Ap star CU Virginis \citep{tri00, tri11, rav10} and of certain types of radio bursts occurring during solar and stellar flares \citep{mel82, dul85, fle98}.

The electron-cyclotron maser emission is highly sensitive to the parameters of plasma, magnetic field, and energetic particles
in its source. However, utilizing the high diagnostic potential of this emission is a difficult task. Although some parameters of the emission source can be estimated using general features of the maser mechanism (e.g., the emission frequency, as a rule, is very close to the electron cyclotron frequency which allows us to find the magnetic field strength), most of the source parameters affect the emission properties in a complicated and nonlinear way. A comprehensive nonlinear simulation of the ECMI requires considering the spatial movement of the waves and particles, which requires a lot of computational resources.

In most works devoted to nonlinear simulation of the ECMI, the spatial movement of waves and particles was simply ignored,
which corresponds to an infinite and homogeneous source (the so-called diffusive limit). Kinetic simulations of relaxation of the loss-cone electron distribution in a diffusive limit were performed by \citet{asc88, asc90, fle00}. Similar simulations for
the horseshoe-like distribution are presented in the paper of \citet{kuz11a} ({\em hereafter Paper I}). A number of simulations in a similar approach (i.e., ignoring the source inhomogeneity and finite size) for different electron distributions were performed using the particle-in-cell technique \citep[e.g.,][]{pri84a, pri84b, pri86a, pri85}. More complicated particle-in-cell simulations considered an inhomogeneous emission source and used specific boundary conditions and/or numerical methods to account for the radiation escape from the source as well as for the particle flow through the emission source region \citep{pri86b, win88, pri89, pri02}.

In this work, we use a different approach. Like in Paper I, the particles and waves are simulated using a kinetic description. The used model does not explicitly include the spatial movement of the particles and waves and their spatial ditributions within the emission source, i.e., the parameters of the particles and waves are treated as averaged over the source volume. The spatial movement of the electromagnetic waves (which results in their escape from the source) is taken into account implicitly by introducing a finite amplification time. In turn, the spatial movement of the electrons is taken into account implicitly by introducing the injection and escape terms in the kinetic equation. Although the considered model is still oversimplified, we believe it is a much better approximation to reality than the diffusive limit. Like in Paper I, we consider the horseshoe-like distribution of the energetic electrons \citep[e.g.,][]{del98, erg00, str01}. The dispersion relation of the electromagnetic waves is assumed to be similar to that in a vacuum. Simulations are performed for different parameter sets corresponding to the plasma parameters measured in the sources of the auroral radio emissions of the Earth and Saturn, and to the expected parameters of the magnetospheres of ultracool dwarfs. Although our model, in principle, can be applied to non-stationary processes, we focus in this work on finding quasi-stationary solutions and analyzing their characteristics.

Generation of terrestrial auroral kilometric radiation in fi\-ni\-te-size low-density plasma cavities was studied by \citet{cal81, cal82, lou96a, lou96b, lou06, bur07}, who focused on the wave dispersion and propagation and the conditions of radiation escape from the cavity, but did not consider the nonlinear wave-particle interaction. The difference of this work is that, firstly, our aim is to investigate the nonlinear wave-particle interaction, while the wave propagation effects are assumed to be incorporated into a single factor -- the wave amplification time. Secondly, we do not consider the possible formation of a discrete spectrum of eigenmodes in a plasma cavity, assuming that the quality factor of such a resonator is too low due to strong energy leakage.

As can be seen from above, we do not consider formation of fine temporal and spectral structures in the dynamic spectra of radio emission. It was recently proposed that these features may be a reflection of small-scale short-lived nonlinear structures (such as electron holes) in the emission source \citep[e.g.,][]{tre11, tre12}. However, simulating the electron-cyclotron maser emission from (or in presence of) electron holes requires a more sophisticated approach and thus is beyond the scope of this paper.

The numerical model used in the simulations is largely similar to that in Paper I; therefore, most formulae and technical details are omitted and only some important facts and expressions are briefly repeated. The used model is described in Section \ref{model}. The simulation results (for a wide range of parameters) are presented in Section \ref{results}. The comparison of the simulation results with the observations is made in Section \ref{comparison}. The conclusions are drawn in Section \ref{conclusion}.

\section{Model}\label{model}
\subsection{Particle dynamics}\label{p_dyn}
In the general case, the time evolution of the electron distribution in a stellar/planetary magnetosphere should be described by the Fokker-Planck equation including the effects of the magnetic field convergence, magnetic-field-aligned electric field, and wave-particle interaction \citep[e.g.,][]{ham90}. However, as said above, solving the exact equation would require enormous computational resources due to necessity to consider explicitly the spatial movement of the particles. In this work, we use a simplified approach where the spatial dependence of the electron distribution function is neglected and the particle movement is treated approximately using the particle fluxes in and out the considered volume. Therefore, the kinetic equation for the  distribution function $f$ can be written in the form
\begin{equation}\label{fp}
\frac{\partial f(\mathbf{p}, t)}{\partial t}=
\left[\frac{\partial f(\mathbf{p}, t)}{\partial t}\right]_{\mathrm{inj}}+
\left[\frac{\partial f(\mathbf{p}, t)}{\partial t}\right]_{\mathrm{esc}}+
\left[\frac{\partial f(\mathbf{p}, t)}{\partial t}\right]_{\mathrm{rel}},
\end{equation}
where $\mathbf{p}$ is the electron momentum and the terms $(\partial f/\partial t)_{\mathrm{inj}}$, $(\partial f/\partial t)_{\mathrm{esc}}$, and $(\partial f/\partial t)_{\mathrm{rel}}$ describe the injection of the energetic electrons into the considered volume, escape of the electrons from this volume, and variation of the distribution function due to the wave-particle interactions, respectively. The effects of the converging magnetic field and the parallel electric field are assumed to be incorporated into the terms $(\partial f/\partial t)_{\mathrm{inj}}$ and $(\partial f/\partial t)_{\mathrm{esc}}$, since the escaped electrons can be then reflected from the magnetic mirror and/or re-accelerated by the electric field, so that they re-appear in the simulation model as the injected electrons (but with a different value of the momentum $\mathbf{p}$).

We assume that the particle injection is time-independent, so that the corresponding term in Eq. (\ref{fp}) can be written in the form
\begin{equation}\label{df_inj}
\left[\frac{\partial f(\mathbf{p}, t)}{\partial t}\right]_{\mathrm{inj}}=
\left(\frac{\partial n_{\mathrm{e}}}{\partial t}\right)_{\mathrm{inj}}\tilde f_{\mathrm{inj}}(\mathbf{p}),
\end{equation}
where $(\partial n_{\mathrm{e}}/\partial t)_{\mathrm{inj}}$ is the constant injection rate and $\tilde f_{\mathrm{inj}}$ is the normalized to unity distribution function of the injected particles.

The particle escape from the radio emission source can be described by the expression
\begin{equation}\label{df_esc}
\left[\frac{\partial f(\mathbf{p}, t)}{\partial t}\right]_{\mathrm{esc}}=
-\frac{f(\mathbf{p}, t)}{\tau_{\mathrm{esc}}(\mathbf{p})},
\end{equation}
where $\tau_{\mathrm{esc}}$ is the characteristic escape time. For the particles with constant (time-independent) velocities, the escape time can be estimated as $\tau_{\mathrm{esc}}\simeq R_z/|\varv_z|$, where $R_z$ is the longitudinal (i.e., along the magnetic field) source size and $\varv_z$ is the longitudinal component of the velocity. However, in situ measurements within the sources of terrestrial and Saturnian auroral radio emissions reveal the presence of magnetic-field-aligned electric fields which accelerate the electrons \citep{erg98, erg00, tre06, lam10, sch11}; as a result, $\varv_z\neq\textrm{const}$ and even the particles with zero initial velocity eventually leave the source region. For simplicity, we assume in this work that the particle escape time is constant and is given by the expression
\begin{equation}\label{tau_esc}
\tau_{\mathrm{esc}}=R_z/\varv_{\mathrm{b}},
\end{equation}
where $\varv_{\mathrm{b}}$ is a typical particle speed. In a quasi-stationary state (when the electron density is constant), the described model is equivalent to the ``recycling'' method used in the particle-in-cell simulations by \citet{win88}.

\subsection{Wave dynamics}\label{w_dyn}
If spatial movement of waves is neglected, the kinetic equation for the waves of mode $\sigma$ can be written in the form
\begin{equation}\label{dWk}
\frac{\partial W_{\mathbf{k}}^{(\sigma)}(\mathbf{k}, t)}{\partial t}=
\gamma^{(\sigma)}(\mathbf{k}, t)W_{\mathbf{k}}^{(\sigma)}(\mathbf{k}, t),
\end{equation}
where $W_{\mathbf{k}}^{(\sigma)}$ is the energy density of the waves in the space of wave vectors, $\mathbf{k}$ is the wave vector, and $\gamma^{(\sigma)}$ is the growth rate (i.e., the above equation describes wave amplification with time due to an instability). The wave-mode index $\sigma$ is hereafter omitted for brevity.

In Paper I, we considered a diffusive limit, when the wave amplification was limited only by nonlinear effects -- relaxation of an unstable electron distribution due to interaction with waves, which resulted in a decrease of the growth rate. In this work, we consider an opposite limit, when the waves are not accumulated in the emission source. Escape of the waves from the amplification region (i.e., from the volume occupied by the electrons with an unstable distribution) naturally limits their intensity. To account for this effect, we assume that (i) the electromagnetic waves are amplified due to the electron-cyclotron maser instability only during a small time interval $\Delta t$ and (ii) the growth rate remains nearly constant during this time interval. As a result, the maximum energy density of the waves can be estimated as
\begin{equation}\label{Wmax}
W_{\mathbf{k}}^{(\max)}(\mathbf{k}, t)=W_0(\mathbf{k})\exp\left[\gamma(\mathbf{k}, t)\Delta t(\mathbf{k}, t)\right],
\end{equation}
where $W_0$ is some initial energy density (which is determined, e.g., by the spontaneous emission processes).

The wave amplification time can be estimated as
\begin{equation}\label{dt_general}
\Delta t\simeq\min\left(\left|\frac{R_z}{\partial r_z/\partial t}\right|, 
\left|\frac{R_{\bot}}{\partial r_{\bot}/\partial t}\right|, 
\left|\frac{\Delta k_z}{\partial k_z/\partial t}\right|, 
\left|\frac{\Delta k_{\bot}}{\partial k_{\bot}/\partial t}\right|\right),
\end{equation}
where $R_{z,\bot}$ are the characteristic sizes of the emission source (along and across the magnetic field, respectively), $\Delta k_{z,\bot}$ are the characteristic sizes of the wave amplification region in the space of wave vectors, and $\partial r_{z,\bot}/\partial t$ and $\partial k_{z,\bot}/\partial t$ are the corresponding movement speeds of a wave packet. Escape of the waves from resonance with energetic electrons due to variation of their wave vector in an inhomogeneous medium can be an important factor affecting, e.g., formation of some types of solar radio bursts \citep{vla02, kuz06, kuz08}. However, we consider in this work the case when the plasma density is very low so that the ``vacuum approximation'' for the wave dispersion can be used (see below) and the wave vector is nearly independent on the medium parameters and thus on the spatial coordinates; as a result, $\partial k_{z,\bot}\to 0$ and the wave escape from resonance in the space of wave vectors is negligible. Also, we consider ring-like or horseshoe-like electron distributions that generate waves mainly in the direction perpendicular to the magnetic field (see, e.g., Paper I). As a result, $|\partial r_z/\partial t|\ll |\partial r_{\bot}/\partial t|$ and only the spatial escape of the emission from the source in the transverse direction needs to be considered. The expression for the amplification time is then reduced to
\begin{equation}\label{dt}
\Delta t=\frac{R_{\bot}}{\varv_{\mathrm{gr}}\sin\theta},
\end{equation}
where $\varv_{\mathrm{gr}}$ is the group speed of the waves and $\theta$ is their propagation angle relative to the magnetic field. We should note that, according to the satellite measurements, the terrestrial auroral kilometric radiation is generated in auroral cavities surrounded by a denser plasma \citep[e.g.,][]{erg98, erg00}. The electromagnetic waves can be reflected from the walls of these cavities and return back thus increasing the time spent in the emission source. However, the reflected waves will have a different (not perpendicular to the magnetic field) direction of propagation and thus will not further participate in resonance with the electrons. Therefore the wave amplification time will still be determined by Eq. (\ref{dt}), which now will characterize the time of wave escape from the amplification region in the space of wave vectors.

The formula (\ref{Wmax}) represents the energy density of the waves leaving the amplification region and thus determines the emission intensity outside the source. By using Eq. (\ref{dWk}), we can find the total energy radiated from unit volume per unit time:
\begin{equation}\label{W_rad}
\left[\frac{\partial W(t)}{\partial t}\right]_{\mathrm{rad}}=
\int\gamma(\mathbf{k}, t)W_{\mathbf{k}}^{(\max)}(\mathbf{k}, t)\,\mathrm{d}^3\mathbf{k}.
\end{equation}

\subsection{Wave-particle interaction}\label{wp_int}
Variation of the electron distribution due to interaction with the electromagnetic waves is described by the equation \citep[e.g.,][]{asc90}
\begin{equation}\label{df_rel}
\left[\frac{\partial f(\mathbf{p}, t)}{\partial t}\right]_{\mathrm{rel}}=
\frac{\partial}{\partial p_i}\left[D_{ij}(\mathbf{p}, t)\frac{\partial f(\mathbf{p}, t)}{\partial p_j}\right],
\end{equation}
where $D_{ij}$ is the diffusion tensor:
\begin{equation}\label{diff}
D_{ij}(\mathbf{p}, t)=
\int\frac{\mathcal{A}(\mathbf{k}, \mathbf{p})}{\omega}k_ik_jW_{\mathbf{k}}(\mathbf{k}, t)\,\mathrm{d}^3\mathbf{k},
\end{equation}
$\omega$ is the wave frequency and $\mathcal{A}$ is the factor describing the efficiency of the wave-particle interaction (it includes the resonance condition and summation over cyclotron harmonics). If several wave modes can exist simultaneously, then the diffusion tensor $D_{ij}$ is the sum of the tensors (\ref{diff}) for the different modes. In turn, the growth rate of the electromagnetic waves due to interaction with the energetic electrons equals \citep[e.g.,][]{asc90}
\begin{equation}\label{incr}
\gamma(\mathbf{k}, t)=
\int\mathcal{A}(\mathbf{k}, \mathbf{p})k_i\frac{\partial f(\mathbf{p}, t)}{\partial p_i}\,\mathrm{d}^3\mathbf{p}.
\end{equation}
The exact expressions for the factor $\mathcal{A}$ and the methods of calculating the growth rate, diffusion tensor coefficients, and diffusion rate are given in many works \citep[e.g.,][]{mel82, asc88, asc90} and summarized in Paper I.

When calculating the diffusion tensor (\ref{diff}), we assume that the wave energy density $W_{\mathbf{k}}$ equals its maximum value $W_{\mathbf{k}}^{(\max)}$ given by Eq. (\ref{Wmax}), since the waves with the highest energy density make the maximum contribution into the relaxation process of an unstable electron distribution. Thus, by using all the assumptions listed in Sections \ref{p_dyn}-\ref{wp_int}, the time derivative of the electron distribution function $\partial f/\partial t$ is reduced to an explicit function of the distribution function itself; in turn, the simulation model is reduced to the differential equation (\ref{fp}). In the below simulations, the electron distribution function $f(\mathbf{p})$ is defined on a regular grid in $(p, \alpha)$ space, where $\alpha$ is the electron pitch angle. Equation (\ref{fp}) is numerically integrated with respect to time using the Gear formulae of fourth order with an adaptive stepsize; the details of the numerical code are given in Paper I.

\subsection{Model parameters}
In situ measurements within the sources of terrestrial and Sa\-tur\-ni\-an auroral radio emissions have revealed that such regions have the following characteristic features \citep[e.g.,][]{tre06, lam10}: (i) the plasma density is very low so that the electron plasma frequency $\omega_{\mathrm{p}}$ is much lower than the electron cyclotron frequency $\omega_{\mathrm{B}}$; (ii) the cold (thermal) electron component is almost absent, so that the energetic electrons with the energy of a few keV dominate. Under these conditions, the dispersion of the electromagnetic waves differs significantly from that in a cold plasma \citep{pri84a, pri84b, leq84a, leq84b, win85, str85, str86, rob86, rob87, leq89, lou96b}. In particular, numerical simulations by \citet{rob86, rob87} have shown that if the plasma parameters satisfy the condition
\begin{equation}\label{vcond}
\varv_{\mathrm{b}}/c\gtrsim\omega_{\mathrm{p}}^2/\omega_{\mathrm{B}}^2
\end{equation}
then the dispersion of the waves (in the case of quasi-transverse propagation) becomes similar to that in a vacuum, i.e., the refraction index $N$ becomes close to unity for both the ordinary and extraordinary waves, the low-frequency cutoff of the X-mode (which is predicted by the cold plasma theory) disappears and the dispersion branches corresponding to the X- and Z-modes of a cold plasma reconnect forming a single branch (with only a small wiggle near the electron cyclotron frequency). Condition (\ref{vcond}) is well satisfied within the sources of terrestrial and Saturnian auroral radio emissions \citep[e.g.,][]{del98, erg00, lam10}; most likely, radio emissions of other magnetized planets and ultracool dwarfs are produced under similar conditions. In addition, observations of the waves within the sources of terrestrial auroral emission have found no indication of the low-frequency X-mode cutoff \citep{erg98}. Therefore, we adopt in this work a vacuum-like dispersion relation, with $N=1$ and $\varv_{\mathrm{gr}}=c$ for both the ordinary and extraordinary modes. Other details of calculating the growth rate and diffusion tensor for such waves are given in Paper I.

\begin{figure*}
\parbox[b]{6.13cm}{\includegraphics[bb=0 0 6.13cm 1.9cm]{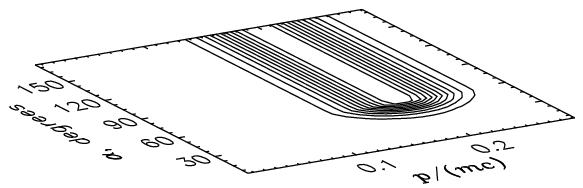}\\
\includegraphics{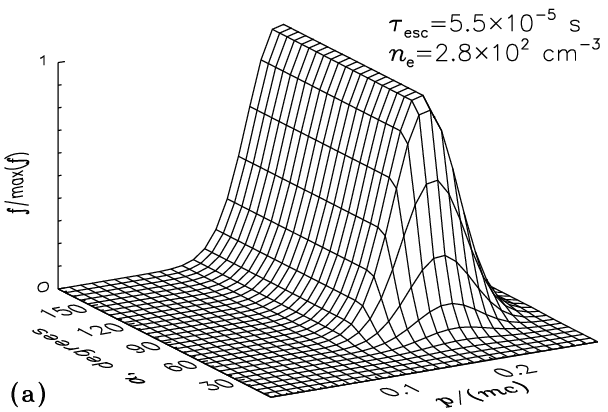}}%
\parbox[b]{6.13cm}{\includegraphics[bb=0 0 6.13cm 1.9cm]{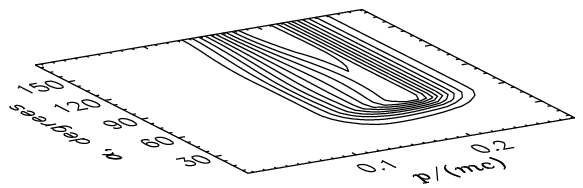}\\
\includegraphics{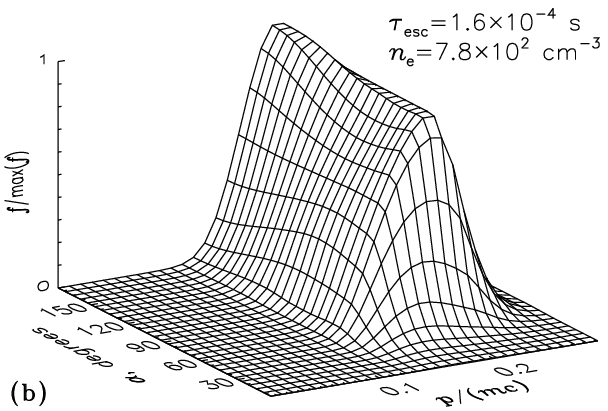}}%
\parbox[b]{6.13cm}{\includegraphics[bb=0 0 6.13cm 1.9cm]{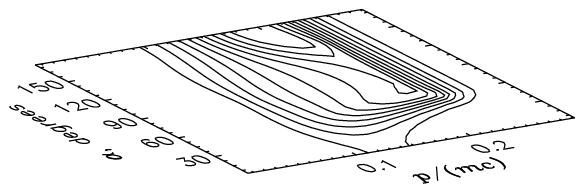}\\
\includegraphics{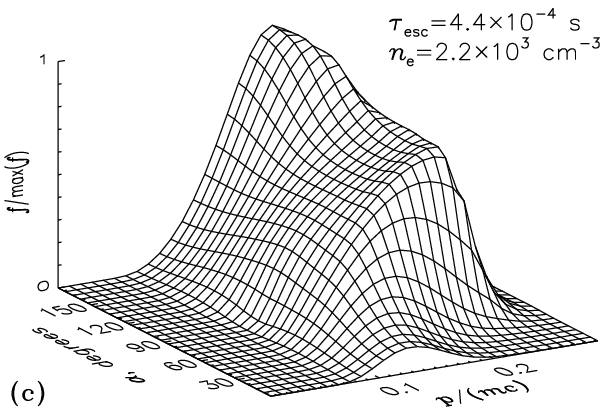}}\\[6pt]
\parbox[b]{6.13cm}{\includegraphics[bb=0 0 6.13cm 1.9cm]{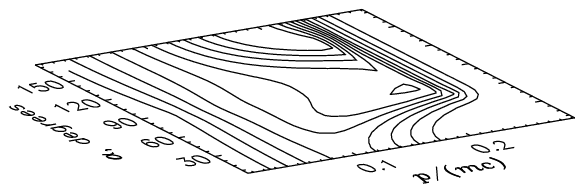}\\
\includegraphics{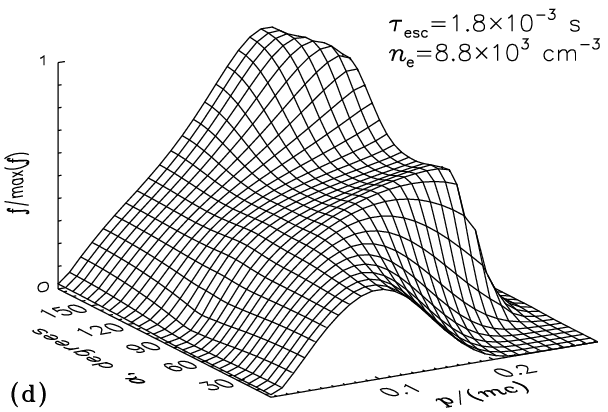}}%
\parbox[b]{6.13cm}{\includegraphics[bb=0 0 6.13cm 1.9cm]{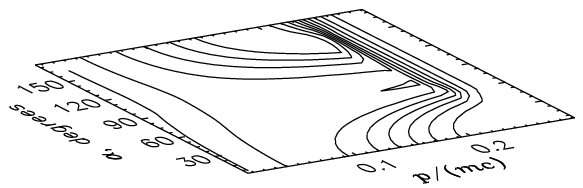}\\
\includegraphics{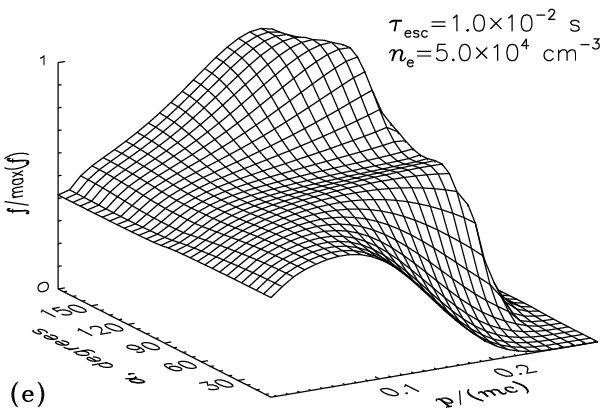}}%
\parbox[b]{6.13cm}{\includegraphics[bb=0 0 6.13cm 1.9cm]{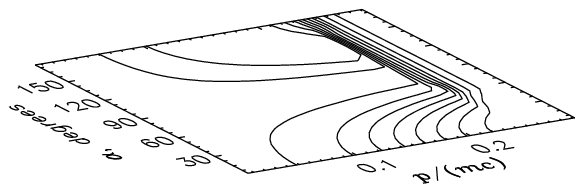}\\
\includegraphics{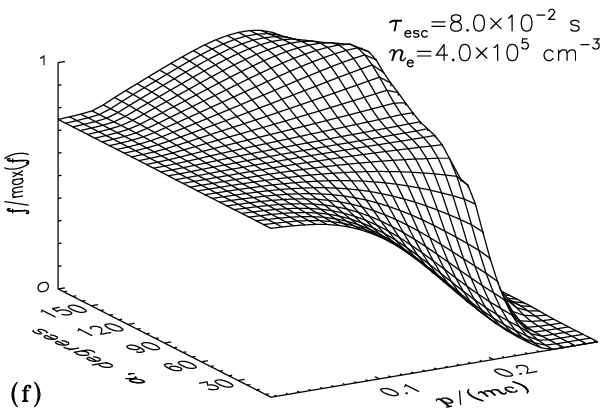}}
\caption{Quasi-stationary electron distributions for the different values of the particle escape time. Each panel represents a normalized distribution function $f/\textrm{max}(f)$ vs. the normalized momentum $p/(m_{\mathrm{e}}c)$ and the pitch-angle $\alpha$. The simulation parameters are given in Section \protect\ref{qss}.}
\label{FigBeam}
\end{figure*}
\begin{figure*}
\parbox[b]{6.13cm}{\includegraphics[bb=0 0 6.13cm 1.9cm]{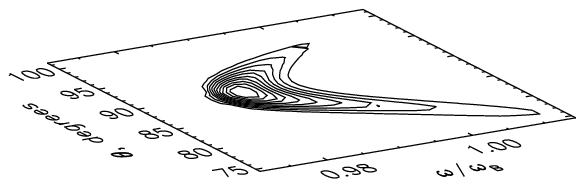}\\
\includegraphics[bb=0 0 6.13cm 3.75cm]{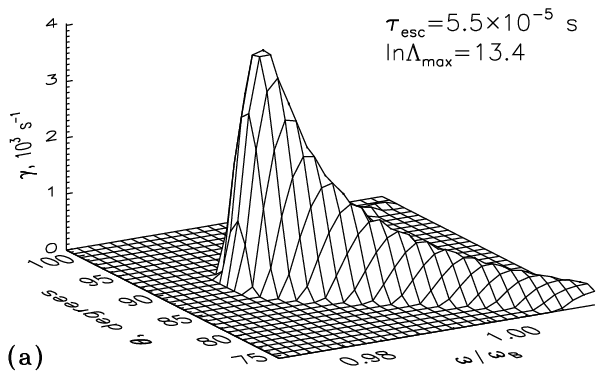}}%
\parbox[b]{6.13cm}{\includegraphics[bb=0 0 6.13cm 1.9cm]{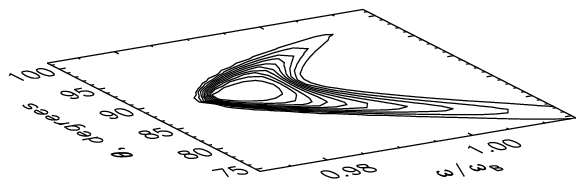}\\
\includegraphics[bb=0 0 6.13cm 3.75cm]{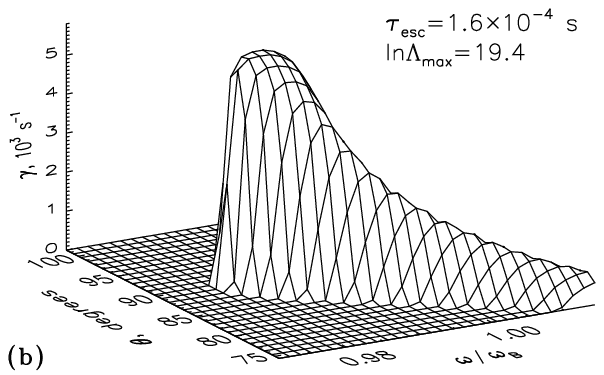}}%
\parbox[b]{6.13cm}{\includegraphics[bb=0 0 6.13cm 1.9cm]{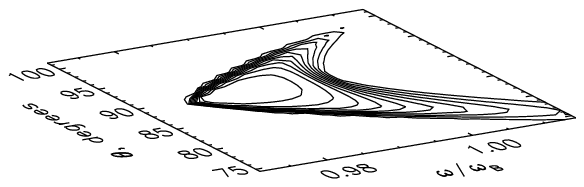}\\
\includegraphics[bb=0 0 6.13cm 3.75cm]{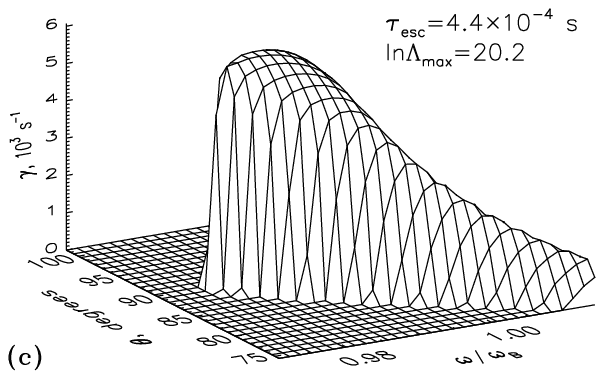}}\\[6pt]
\parbox[b]{6.13cm}{\includegraphics[bb=0 0 6.13cm 1.9cm]{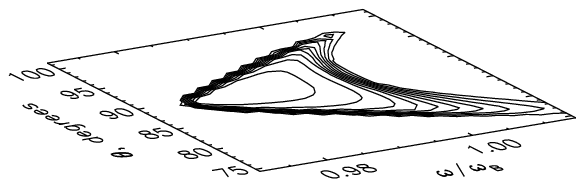}\\
\includegraphics[bb=0 0 6.13cm 3.75cm]{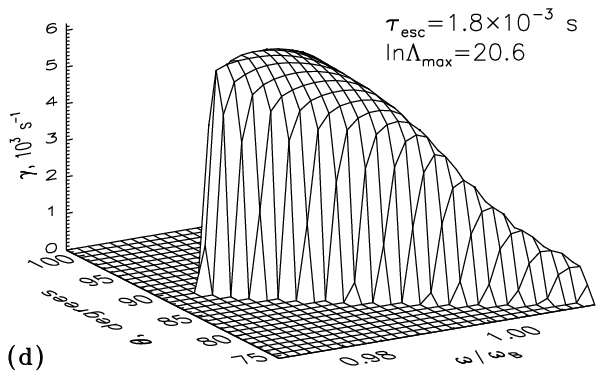}}%
\parbox[b]{6.13cm}{\includegraphics[bb=0 0 6.13cm 1.9cm]{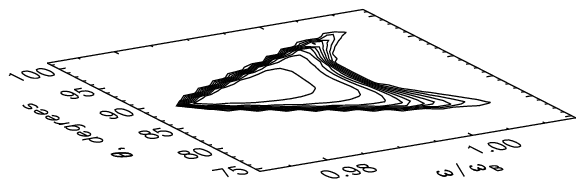}\\
\includegraphics[bb=0 0 6.13cm 3.75cm]{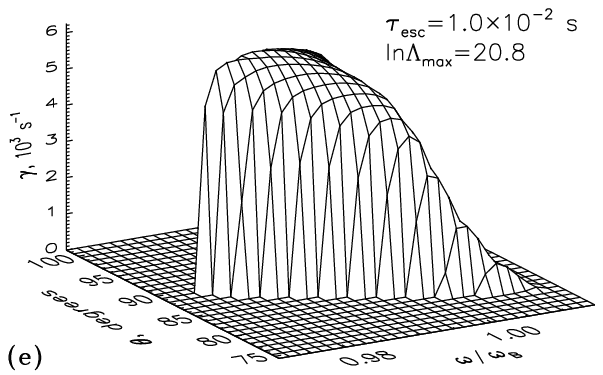}}%
\parbox[b]{6.13cm}{\includegraphics[bb=0 0 6.13cm 1.9cm]{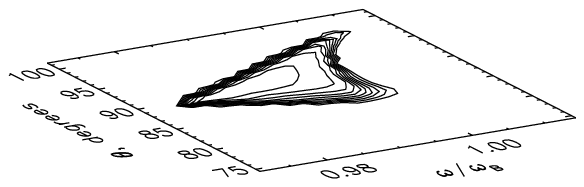}\\
\includegraphics[bb=0 0 6.13cm 3.75cm]{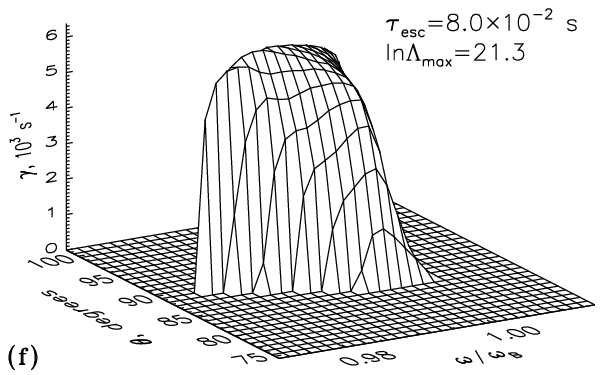}}
\caption{Quasi-stationary growth rates of the extraordinary waves for the different values of the particle escape time. Each panel represents a growth rate $\gamma$ vs. the normalized frequency $\omega/\omega_{\mathrm{B}}$ and the propagation direction $\theta$. The simulation parameters are given in Section \protect\ref{qss}.}
\label{FigIncr}
\end{figure*}

We assume in this work that the injected energetic electrons have a horseshoe-like distribution \citep{del98, erg00, str01}, which is modeled by the following expression (the same as in Paper I):
\begin{equation}\label{f_inj}
\tilde f_{\mathrm{inj}}(\mathbf{p})=A\exp\left[-\frac{(p-p_{\mathrm{b}})^2}{\Delta p_{\mathrm{b}}^2}\right]
\left\{\begin{array}{ll}
1, & \mu\le\mu_{\mathrm{c}},\\
\displaystyle\exp\left[-\frac{(\mu-\mu_{\mathrm{c}})^2}{\Delta\mu_{\mathrm{c}}^2}\right], & \mu>\mu_{\mathrm{c}},
\end{array}\right.
\end{equation}
where $p_{\mathrm{b}}$ and $\Delta p_{\mathrm{b}}$ are the typical momentum of the injected electrons and their dispersion in momentum, respectively, $\mu=\cos\alpha$, $\mu_{\mathrm{c}}=\cos\alpha_{\mathrm{c}}$, $\alpha_{\mathrm{c}}$ and $\Delta\mu_{\mathrm{c}}$ are the loss-cone boundary and the boundary width, respectively, and $A$ is the normalization factor. The electrons with $0\le\alpha<\pi/2$ move upwards (i.e., toward a decreasing magnetic field), and the electrons with $\pi/2<\alpha\le\pi$ move downwards (toward an increasing magnetic field). If $\alpha_{\mathrm{c}}=0$, we obtain an isotropic ring-like distribution.

The initial wave energy density $W_0$ depends on several factors, including the spontaneous radiation and the emission coming from beyond the considered region. We characterize this energy density by the corresponding effective temperature $T_0$, i.e.
\begin{equation}\label{W0}
W_0=\frac{k_{\mathrm{B}}T_0}{(2\pi)^3}=\mathrm{const},
\end{equation}
where $k_{\mathrm{B}}$ is the Boltzmann constant.

When solving the differential equation (\ref{fp}), we are interested in finding a quasi-stationary state with $\partial f/\partial t\to 0$. Therefore, the initial electron distribution function (at $t=0$) is unimportant. We either assumed that the initial distribution function was equal to zero, or used a result of another simulation as the initial condition for the next simulation run, in order to provide faster convergence. Simulations were stopped when the variation rate of the distribution function fell below a certain threshold ($10^{-3}-10^{-2}$ of the injection rate).

\section{Simulation results}\label{results}
\subsection{Electron distributions and emission characteristics in a quasi-stationary state}\label{qss}
It is easy to show analytically (and is confirmed by our numerical simulations) that for the above described model, with an increasing time, the electron distribution asymptotically approaches a quasi-stationary state with $\partial f/\partial t\to 0$ regardless of the initial distribution, with $\tau_{\mathrm{esc}}$ being the characteristic timescale of this process. The same conclusion can be made for the emission intensity and spectrum, since they are determined by the electron distribution at a given time. Thus if the characteristic variation timescales of the observed radio emission far exceed the particle escape time $\tau_{\mathrm{esc}}$ (which implies that the source parameters are stable at those timescales), we can expect that the electron distribution in the emission source has reached a quasi-stationary state or is very close to it. Radio emissions of the planets and ultracool dwarfs, as a rule, satisfy this condition (see Section \ref{comparison}). Therefore we focus in this work on the electron distributions and the emission intensities and spectra in a quasi-stationary state.

\begin{figure}
\centerline{\includegraphics{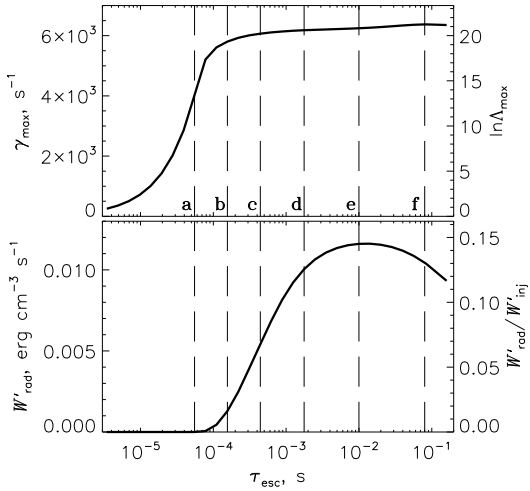}}
\caption{Maximum growth rate (top) and total emission intensity (bottom) of the extraordinary waves in a quasi-stationary state vs. the particle escape time. The vertical lines correspond to the values of $\tau_{\mathrm{esc}}$ shown in Figs. \protect\ref{FigBeam}-\protect\ref{FigIncr}. The simulation parameters are given in Section \protect\ref{qss}.}
\label{FigSummary}
\end{figure}

Firstly, we consider the relative effect of the two factors affecting the electron distribution: particle escape from the radio emission source and relaxation of the electron distribution due to wave-particle interaction. We use the following simulation parameters (that were chosen so as to reproduce the observed characteristics of radio emission from ultracool dwarfs, see Section \ref{comparison} for details): the electron cyclotron frequency $\nu_{\mathrm{B}}=4.5$ GHz, the transverse size of the emission source $R_{\bot}=1000$ km, the initial effective temperature of the waves $T_0=10^6$ K, the typical energy of the injected electrons $E_{\mathrm{b}}=10$ keV, the relative dispersion of the electrons in momentum $\Delta p_{\mathrm{b}}/p_{\mathrm{b}}=0.2$, the loss-cone boundary $\alpha_{\mathrm{c}}=60^{\circ}$, the loss-cone boundary width $\Delta\mu_{\mathrm{c}}=0.2$, and the electron injection rate $(\partial n_{\mathrm{e}}/\partial t)_{\mathrm{inj}}=5\times 10^6$ $\textrm{cm}^{-3}$ $\textrm{s}^{-1}$. The particle escape time $\tau_{\mathrm{esc}}$ is variable. Since the extraordinary mode has been found to be strongly dominant under the above conditions (see Section \ref{modes}), we consider only this mode in most of our simulations. Figure \ref{FigBeam} shows the quasi-stationary electron distributions corresponding to the different values of the particle escape time. Note that the distribution function in each panel of Fig. \ref{FigBeam} is normalized to its maximum value, while the actual values of the distribution function increase with $\tau_{\mathrm{esc}}$ which is indicated by the increasing total electron density $n_{\mathrm{e}}$. Figure \ref{FigIncr} shows the corresponding growth rates of the extraordinary waves. Finally, Figure \ref{FigSummary} summarizes the simulation results for the different escape times and shows the maximum growth rate and the total radiation energy flux from unit volume calculated using Eq. (\ref{W_rad}). The parameter $\Lambda$ in Figs. \ref{FigIncr}-\ref{FigSummary} is the wave amplification factor that is defined by $\Lambda=W_{\mathbf{k}}/W_0$; according to Eq. (\ref{Wmax}), $\ln\Lambda=\gamma\Delta t$.

\begin{figure}
\centerline{\includegraphics{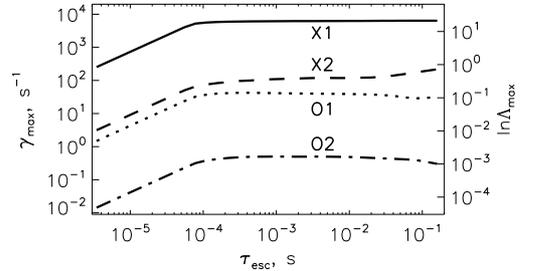}}
\caption{Maximum growth rate of different wave modes in a quasi-stationary state vs. the particle escape time. The simulation parameters are given in Section \protect\ref{qss}.}
\label{FigModes}
\end{figure}

\begin{figure*}
\centerline{\includegraphics{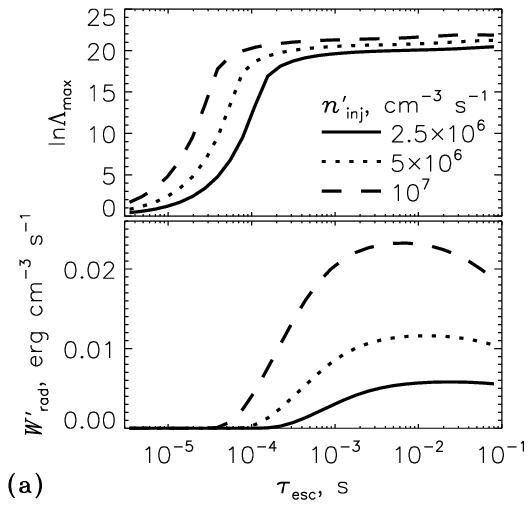} 
\includegraphics{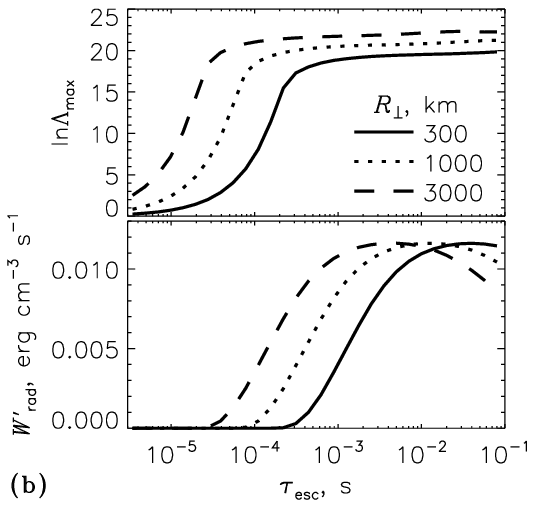} 
\includegraphics{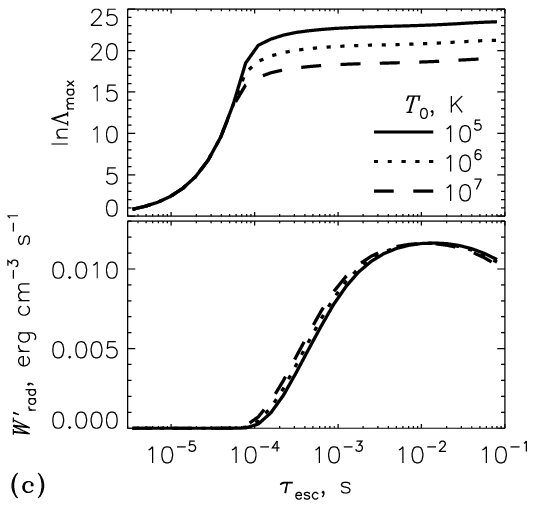}}
\centerline{\includegraphics{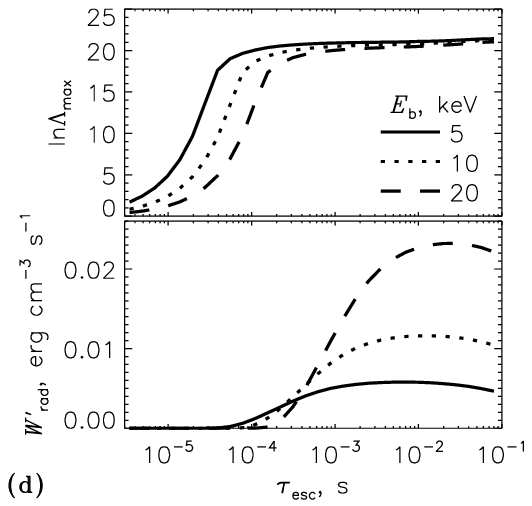} 
\includegraphics{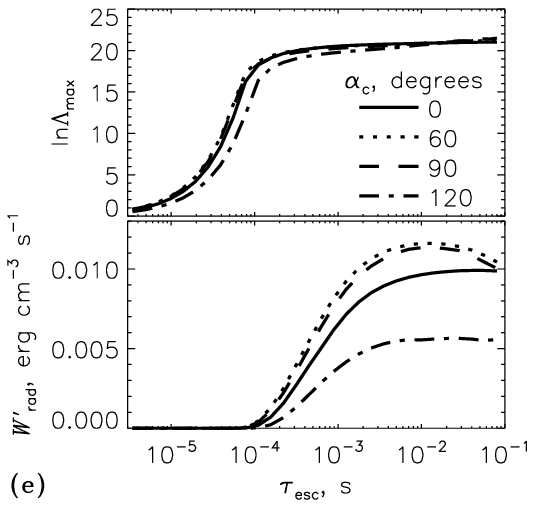} 
\includegraphics{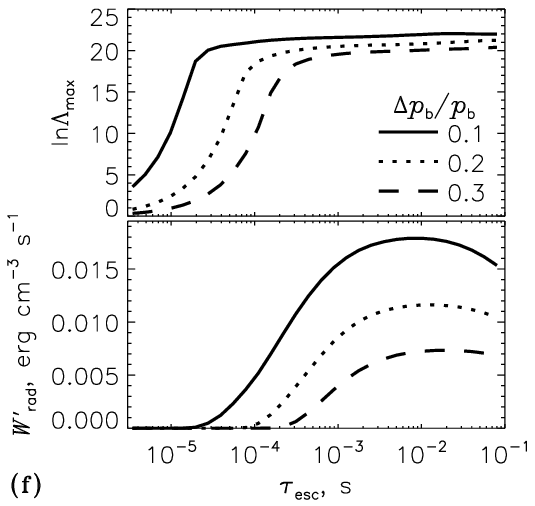}}
\caption{Maximum growth rates and total emission intensities of the extraordinary waves in a quasi-stationary state vs. the particle escape time for the different parameter sets. In each panel, most simulation parameters are the same as in Section \protect\ref{qss}, and one parameter (whose values are specified in the panel) is variable.}
\label{FigParams}
\end{figure*}

To avoid a possible confusion, we highlight here that Fig. \ref{FigBeam} does not represent a temporal evolution of an electron distribution. Instead, each panel in Fig. \ref{FigBeam} represents a final (quasi-stationary) stage of such evolution. These final distributions correspond to different operating conditions of the electron-cyclotron maser, namely, to different longitudinal source sizes $R_z$ which results in different values of the particle escape time $\tau_{\mathrm{esc}}$. Accordingly, Fig. \ref{FigIncr} represents the final (quasi-stationary) growth rates, while Figs. \ref{FigSummary}-\ref{FigParams} plot some characteristic parameters of the quasi-stationary solutions vs. the emission source parameter $\tau_{\mathrm{esc}}$.

Figures \ref{FigBeam}a and \ref{FigIncr}a correspond to a case when the timescale $\tau_{\mathrm{esc}}$ is very small and thus  the particle escape from the emission source region is very fast. In this case, the electron density (and, consequently, the wave growth rate) cannot reach a level sufficient to provide a significant wave amplification. As a result, relaxation of the electron distribution due to interaction with the waves is negligible and the electron distribution in a quasi-stationary state does not differ from the distribution of the injected electrons $\tilde f_{\mathrm{inj}}$. The corresponding electron density equals
\begin{equation}\label{n_inf}
n_{\infty}=\left(\frac{\partial n_{\mathrm{e}}}{\partial t}\right)_{\mathrm{inj}}\tau_{\mathrm{esc}}.
\end{equation}
The growth rate has a sharp peak at $\theta\simeq 90^{\circ}$ and $\omega/\omega_{\mathrm{B}}\simeq 0.985$ that corresponds to the relativistic cyclotron frequency of the electrons with the energy of about 10 keV. The emission intensity is very low (see Fig. \ref{FigSummary}).

The quasi-stationary growth rate of the waves increases linearly with $\tau_{\mathrm{esc}}$, until the corresponding wave energy density exceeds a certain threshold so that the wave-particle interaction begins to affect the electron distribution. Under the considered conditions, this happens at $\ln\Lambda\gtrsim 19$. Figure \ref{FigBeam}b shows an example of a weakly-relaxed quasi-stationary state when the electron distribution is only slightly distorted in comparison with the distribution of the injected electrons. In turn, the relaxation of the electron distribution reduces the wave growth rate, so that the growth rate peak broadens and flattens (see Fig. \ref{FigIncr}b). Note that the wave energy density depends exponentially on the growth rate, and therefore only the waves in a very narrow spectral ($\Delta\omega/\omega\simeq 0.01$) and angular ($\Delta\theta\simeq 3^{\circ}$) range actually make a significant contribution into the total emission intensity as well as into the relaxation of the electron distribution. Although the fraction of the particle energy flux that is transferred to waves is relatively low (about 1\% for the state shown in Figs. \ref{FigBeam}b and \ref{FigIncr}b), this conversion efficiency can be sufficient to provide, e.g., the observed intensity of terrestrial and Saturnian auroral kilometric radio emissions \citep{gur74, ben79, kur05, lam11}.

For even larger values of the timescale $\tau_{\mathrm{esc}}$ (Figs. \ref{FigBeam}c-\ref{FigBeam}d), the electrons interact with the waves for a longer time and thus drift further in velocity space (towards lower velocities) before they eventually leave the radio emission source. As a result, the particle dispersion in velocity increases with $\tau_{\mathrm{esc}}$; for the case shown in Fig. \ref{FigBeam}d, the empty space inside the ``horseshoe'' is almost filled. The maximum growth rate remains nearly constant, but the flattened region around the peak of the growth rate broadens (Figs. \ref{FigIncr}c-\ref{FigIncr}d); therefore, the total emission intensity gradually increases with $\tau_{\mathrm{esc}}$.


Figures \ref{FigBeam}e and \ref{FigIncr}e correspond to the value of the timescale $\tau_{\mathrm{esc}}$ that provides the maximum emission intensity; the conversion efficiency of the particle energy flux into waves in the quasi-stationary state reaches 14.5\%. One can see that the empty space around $\varv=0$ is now completely filled by the electrons; the quasi-stationary electron distribution does not look like a ``horseshoe'' but rather represents a combination of a nearly flat plateau and a downgoing electron beam (with energy of $\sim 10$ keV and the pitch-angle dispersion of $\sim 45^{\circ}$). With a further increase of $\tau_{\mathrm{esc}}$, the electron distribution does not change qualitatively -- the plateau level steadily rises while the beam-like component remains nearly the same (see Fig. \ref{FigBeam}f). The peak of the wave growth rate becomes narrower (Fig. \ref{FigIncr}f) and the radio emission characteristics become similar to those for a beam-driven electron-cyclotron instability, i.e., the emission becomes directed slightly downwards (with the maximum at $\theta\simeq 90^{\circ}-93^{\circ}$). Although the maximum growth rate slightly increases with $\tau_{\mathrm{esc}}$, the total emission intensity steadily decreases but remains sufficiently high (with an energy conversion efficiency of $\gtrsim 10\%$).

Equation (\ref{n_inf}) for the electron density in a quasi-stationary state $n_{\infty}$ was obtained under the assumption that the wave-particle interaction is negligible. If such interaction is important, the quasi-stationary electron density slightly exceeds $n_{\infty}$ (but by no more than $10^{-3}$).

Note that the values of the particle escape time used in this Section were chosen rather arbitrarily. We have found that different ratios of the particle injection and escape rates, together with the wave-particle interaction, yield qualitatively different electron distributions and radio emission characteristics. Consequently, the actual characteristics of the particles and radio emission in a planetary or stellar magnetosphere will be determined by the local conditions (such as the emission source dimensions, electron acceleration rate, etc.). In particular, the electron distributions observed in terrestrial or Saturnian magnetosphere seem to be weakly or moderately relaxed like those shown in Figs. \ref{FigBeam}b-\ref{FigBeam}c, while the electron distribution in the magnetospheres of ultracool dwarfs are expected to be strongly relaxed like that shown in Fig. \ref{FigBeam}f (see Section \ref{comparison} for details).

\subsection{Relative contribution of different wave modes}\label{modes}
In a low-density plasma (with $\omega_{\mathrm{p}}/\omega_{\mathrm{B}}\ll 1$), the fundamental extraordinary mode is expected to be the dominant mode of the electron-cyclotron maser \citep[e.g.,][]{wu79, mel82, sha84, win85, fle94, kuz11a}. This conclusion is generally supported by the observations of the planetary auroral radio emissions \citep{zar98, tre06, erg00, lam11}. In order to investigate a possible contribution of other wave modes, we performed a simulation considering four modes simultaneously: fundamental and harmonic extraordinary modes (X1 and X2), and fundamental and harmonic ordinary modes (O1 and O2). The simulation parameters were the same as in the previous Section.

Figure \ref{FigModes} demonstrates the quasi-stationary growth rates of the considered wave modes corresponding to the different values of the particle escape time $\tau_{\mathrm{esc}}$. One can see that the ratios of the growth rates are nearly independent on $\tau_{\mathrm{esc}}$, i.e., they are nearly the same for both the weakly and strongly relaxed electron distributions. Although the maximum growth rate of the X2-mode somewhat increases at large $\tau_{\mathrm{esc}}$, it always remains much lower (by more than an order of magnitude) than that of the X1-mode. Since the emission intensity depends exponentially on the growth rate, the X2-mode intensity is lower than that of the X1-mode by a factor of $\sim 10^9$. The growth rates and hence intensities of the O1- and O2-modes are even lower. Thus we conclude that the X1-mode is strongly dominant under the considered conditions. The intensities of other modes are negligible, and therefore they do not make a contribution into both the escaping radiation and the relaxation of the electron distribution.

\subsection{Effects of varying the model parameters}\label{params}
In Section \ref{qss}, we considered the particle and emission characteristics for the different values of the particle escape time from the emission source, i.e., for the different longitudinal sizes of the source. Now we consider the effect of varying the other model parameters. Figure \ref{FigParams} is a collection of plots similar to those in Fig. \ref{FigSummary}, but corresponding to the different parameter sets. Each panel of Fig. \ref{FigParams} demonstrates the effect of varying a particular model parameter, while the other parameters are assumed to be the same as in Section \ref{qss}.

Figure \ref{FigParams}a demonstrates the effect of varying the particle injection rate $(\partial n_{\mathrm{e}}/\partial t)_{\mathrm{inj}}$. As expected, an increase of this parameter results in an increase of the emission intensity; the quasi-stationary growth rate of the waves also somewhat increases. In addition, a higher particle injection rate allows the emission to be efficiently produced in smaller sources (i.e., with a lower particle escape timescale $\tau_{\mathrm{esc}}$).

\begin{table*}
\caption{Parameters of the considered sources of electron-cyclotron maser radiation.}
\label{TableObs}
\renewcommand{\tabcolsep}{4.7pt}
\centering
\begin{tabular}{lccccccccc}
\hline\hline
\rule[-5pt]{0pt}{15pt}Object & $\nu_{\mathrm{B}}$ & $L$ & $R/R_0$ & $I_{\mathrm{obs}}$ & $d$ & $R_{\bot}$, km & $n_{\mathrm{e}}$, $\textrm{cm}^{-3}$ & $E_{\mathrm{b}}$, keV & $W'_{\mathrm{obs}}$, erg $\textrm{cm}^{-3}$ $\textrm{s}^{-1}$\\
\hline
\rule{0pt}{10pt}Earth (\#1)$^a$ & 400 kHz & $\sim 20$ & $\sim 1.61$ & $\gtrsim 5\times 10^{-21}$ W $\textrm{m}^{-2}$ $\textrm{Hz}^{-1}$ & 1 AU & $\lesssim 100$ & $\lesssim 1$ & $\sim 1-10$ & $\gtrsim 2.9\times 10^{-9}$\\
\rule{0pt}{10pt}Earth (\#2)$^b$ & $\sim 400$ kHz & $\sim 20$ & $\sim 1.61$ & $\lesssim 10^{-8}$ W $\textrm{m}^{-2}$ $\textrm{Hz}^{-1}$ & $\sim 1100$ km & $\sim 350$ & 0.2-0.5 & $\lesssim 10$ & $\lesssim 2.5\times 10^{-8}$\\
\rule{0pt}{10pt}Saturn$^c$ & 10 kHz & $\sim 20$ & $\sim 4.75$ & $\lesssim 10^{-18}$ W $\textrm{m}^{-2}$ $\textrm{Hz}^{-1}$ & 1 AU & $\sim 1000$ & 0.003-0.01 & $\lesssim 10$ & $\lesssim 5.2\times 10^{-12}$\\
\rule{0pt}{10pt}TVLM 513$^d$ & 4.5 GHz & 2.15 & 1.07 & $\lesssim 20$ mJy & 10.5 pc & $\sim 1000$ (?) & ? & ? & $\lesssim 8.2\times 10^{-3}$\\
\hline
\end{tabular}
\tablebib{$^{(a)}$~\citet{zar98, tre06}; $^{(b)}$~\citet{erg00}; $^{(c)}$~\citet{lam10, lam11}; $^{(d)}$~\citet{kuz11b}.}
\end{table*}
\begin{table*}
\caption{Simulation parameters and the resulting radio emission characteristics in a quasi-stationary state for the different radiation sources.}
\label{TableModel}
\centering
\begin{tabular}{lccccccccccc}
\hline\hline
\rule[-5pt]{0pt}{15pt}Object & $\nu_{\mathrm{B}}$ & $R_z$, km & $R_{\bot}$, km & $E_{\mathrm{b}}$, keV & $\alpha_{\mathrm{c}}$ & $\Delta p_{\mathrm{b}}/p_{\mathrm{b}}$ & $n_{\mathrm{e}}$, $\textrm{cm}^{-3}$ & $\tau_{\mathrm{esc}}$, s & $W'_{\mathrm{rad}}$, erg $\textrm{cm}^{-3}$ $\textrm{s}^{-1}$ & $W'_{\mathrm{rad}}/W'_{\mathrm{inj}}$ & $\Delta\theta_{\mathrm{rad}}$\\
\hline
\rule{0pt}{10pt}Earth (\#1) & 400 kHz & 2100 & 100 & 10 & $45^{\circ}$ & 0.1 & 0.5 & 0.036 & $7.9\times 10^{-9}$ & 0.036 & $\sim 1.7^{\circ}$\\
\rule{0pt}{10pt}Earth (\#2) & 400 kHz & 2100 & 350 & 10 & $45^{\circ}$ & 0.1 & 0.5 & 0.036 & $2.4\times 10^{-8}$ & 0.11 & $\sim 2.0^{\circ}$\\
\rule{0pt}{10pt}Saturn & 10 kHz & 200\,000 & 1000 & 10 & $5^{\circ}$ & 0.1 & 0.01 & 3.4 & $6.1\times 10^{-12}$ & 0.13 & $\sim 1.9^{\circ}$\\
\rule{0pt}{10pt}TVLM 513 & 4.5 GHz & 4900 & 1000 & 10 & $60^{\circ}$ & 0.2 & $4.2\times 10^5$ & 0.084 & $1.0\times 10^{-2}$ & 0.13 & $\sim 2.8^{\circ}$\\
\hline
\end{tabular}
\end{table*}

Figure \ref{FigParams}b demonstrates the effect of varying the transverse size of the emission source $R_{\bot}$ and hence the wave amplification time $\Delta t$. If the particle escape timescale is relatively low, an increase of the transverse source size makes the wave amplification more efficient and thus increases the emission intensity significantly. In contrast, if $\tau_{\mathrm{esc}}$ is large, a larger transverse source size means a stronger relaxation of the electron distribution; this may result in a slight decrease of the emission intensity. Note that we talk here about the average emission energy flux from unit volume $(\partial W/\partial t)_{\mathrm{rad}}$; the emission intensity from the entire source volume is proportional to $(\partial W/\partial t)_{\mathrm{rad}}R_{\bot}^2$ and therefore always increases with $R_{\bot}$. Also note that the actual parameter affecting the wave energy density is the amplification time $\Delta t$. If the wave dispersion differs from the ``vacuum approximation'' used in this work, the amplification time may be also different. However, the effect of increasing or decreasing the wave amplification time will be the same as that shown in Fig. \ref{FigParams}b.

Figure \ref{FigParams}c demonstrates the effect of varying the initial effective temperature of the waves $T_0$. One can see that this parameter affects the quasi-stationary growth rate of the waves provided that the wave energy density has reached the relaxation threshold (i.e., it is sufficiently high to affect the electron distribution). A higher initial intensity means that the waves need less amplification to reach the mentioned threshold. On the other hand, the emission intensity itself is almost independent on $T_0$. Therefore we conclude that the initial effective temperature of the waves may be chosen rather arbitrarily; in most simulations, we set it to $10^6$ K.

Figure \ref{FigParams}d shows the simulation results for the different energies of the injected electrons $E_{\mathrm{b}}$. Since the particle injection rate $(\partial n_{\mathrm{e}}/\partial t)_{\mathrm{inj}}$ is assumed to be constant, an increase of the energy $E_{\mathrm{b}}$ corresponds to an increase of the energy injection rate. In turn, this results in an increase of the emission intensity. The maximum conversion efficiency of the particle energy flux into waves is about $14.5\%$ for all considered values of $E_{\mathrm{b}}$.

Figure \ref{FigParams}e shows the simulation results for the different loss-cone angles of the injected electrons $\alpha_{\mathrm{c}}$. If the injected electrons have the isotropic ring-like distribution ($\alpha_{\mathrm{c}}=0$), the emission intensity does not exhibit a decrease at large values of $\tau_{\mathrm{esc}}$; the conversion efficiency of the particle energy flux into waves reaches $12.4\%$. The highest emission intensity (with the conversion efficiency of up to $\sim 14.5\%$) is reached for the horseshoe-like distributions with the loss-cone angles of about $60^{\circ}-90^{\circ}$. The values of $\alpha_{\mathrm{c}}>90^{\circ}$ correspond to the beam-like distributions. As the injected electron beam becomes more collimated (i.e., with an increasing $\alpha_{\mathrm{c}}$), the efficiency of the electron-cyclotron maser and hence the emission intensity rapidly decrease; for $\alpha_{\mathrm{c}}=120^{\circ}$, the maximum conversion efficiency of the particle energy flux into waves is about $7.0\%$.

Finally, Fig. \ref{FigParams}f demonstrates the effect of varying the dispersion of the injected electrons in momentum $\Delta p_{\mathrm{b}}$. The nearly monoenergetic electron beams provide higher growth rate of the waves and possess more free energy than the beams with a larger dispersion in energy. Therefore, a decrease of the parameter $\Delta p_{\mathrm{b}}$ increases the efficiency of the electron-cyclotron maser. One can see in Fig. \ref{FigParams}f that in the models with a smaller $\Delta p_{\mathrm{b}}$, the emission can be efficiently produced in smaller sources, with a lower particle escape timescale $\tau_{\mathrm{esc}}$ and hence by the electron beams with a lower density. The maximum conversion efficiency of the particle energy flux into waves increases with a decreasing $\Delta p_{\mathrm{b}}$ and reaches $22.4\%$, $14.5\%$, and $9.2\%$ for $\Delta p_{\mathrm{b}}/p_{\mathrm{b}}=0.1$, $0.2$, and $0.3$, respectively.

\section{Comparison with observations}\label{comparison}
In this Section, we compare the results of our numerical simulations with observations. Namely, we consider the auroral kilometric radio emissions of the Earth and Saturn (for which in situ observations within the source regions are available), and the radio emission of the ultracool dwarf TVLM 513-46546 ({\em hereafter TVLM 513}). Table \ref{TableObs} summarizes the main characteristics of the considered emission sources: the electron cyclotron frequency $\nu_{\mathrm{B}}$, the $L$-shell number of the magnetic field line where the source is located, the distance from the planet/star center $R$ (relative to the planet or star radius $R_0$), the spectral intensity of the emission $I_{\mathrm{obs}}$ at the distance $d$ from the source; the transverse size $R_{\bot}$, the electron density $n_{\mathrm{e}}$ and energy $E_{\mathrm{b}}$ (where known), and the radiation energy flux per unit volume $(\partial W/\partial t)_{\mathrm{obs}}$. The latter parameter was estimated in the following way: since the emission frequency $\nu$ almost coincides with the local electron cyclotron frequency, the observed emission in the frequency range $\mathrm{d}\nu$ is produced in the volume $\mathrm{d}V\simeq R_{\bot}^2\mathrm{d}z$, where the height range $\mathrm{d}z=(\mathrm{d}\nu/\nu)L_{\mathrm{B}}$ and $L_{\mathrm{B}}$ is the magnetic field inhomogeneity scale along the magnetic field line (for the considered sources, $L_{\mathrm{B}}\simeq R/3$). Therefore, the average radiation energy flux per unit volume can be estimated as
\begin{equation}\label{Wobs}
\left(\frac{\partial W}{\partial t}\right)_{\mathrm{obs}}\simeq
I_{\mathrm{obs}}\Delta\Omega\frac{d^2}{R_{\bot}^2}\frac{\nu}{L_{\mathrm{B}}},
\end{equation}
where $\Delta\Omega$ is the emission solid angle. We used the value of $\Delta\Omega\simeq 0.22$ which corresponds to the emission perpendicular to the magnetic field with the beam width of $\Delta\theta_{\mathrm{rad}}\simeq 2^{\circ}$.

Table \ref{TableModel} lists the parameters used in the simulations, as well as the simulation results. The effective longitudinal source size $R_z$ was calculated as the distance along the magnetic field line between the given point (i.e., where the emission with the given frequency is produced) and the lower (high-frequency) boundary of the radio-emitting region. For the auroral radio emissions of the Earth and Saturn, this boundary corresponds to the electron cyclotron frequency of about 800 kHz \citep{zar98, lam11}; for TVLM 513, the high-frequency boundary of the radio spectrum is unknown and therefore we assumed that the radio-emitting region extends down to the stellar photosphere (i.e., to $R/R_0\simeq 1$). The characteristic particle escape time from the emission source region $\tau_{\mathrm{esc}}$ was calculated using Eq. (\ref{tau_esc}), and the loss-cone boundary $\alpha_{\mathrm{c}}$ was calculated using the transverse adiabatic invariant:
\begin{equation}\label{alpha_c}
\sin^2\alpha_{\mathrm{c}}=\nu_{\mathrm{B}}/\nu_{\mathrm{Bmax}},
\end{equation}
where $\nu_{\mathrm{Bmax}}$ is the electron cyclotron frequency at the above-mentioned lower boundary of the radio-emitting region. The electron injection rate was chosen so as to obtain the specified particle density $n_{\mathrm{e}}$ in a quasi-stationary state. In all simulations, we assumed that the initial effective temperature of the electromagnetic waves $T_0=10^6$ K. The simulation results presented include the radiation energy flux from unit volume $(\partial W/\partial t)_{\mathrm{rad}}$ in a quasi-stationary state, the corresponding conversion efficiency of the particle energy flux into waves $(\partial W/\partial t)_{\mathrm{rad}}/(\partial W/\partial t)_{\mathrm{inj}}$, and the emission beam width (at the 1/e level) $\Delta\theta_{\mathrm{rad}}$.

\begin{figure}
\includegraphics{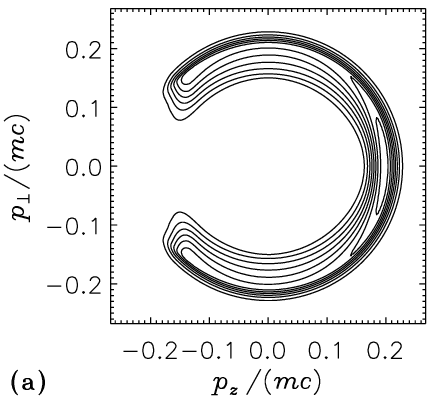}%
\includegraphics{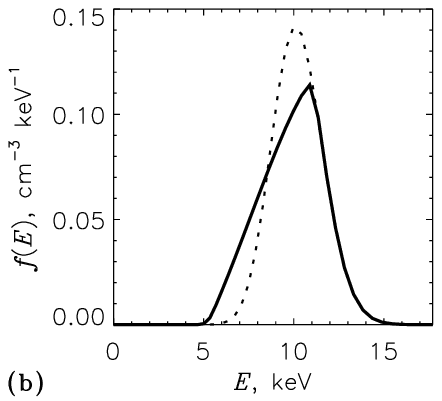}
\caption{Simulated quasi-stationary electron distribution in the magnetosphere of the Earth (model \#1, $R_{\bot}=100$ km). a) 2D distribution function $f(p_z, p_{\bot})$. The $p_z$ axis is reversed according to the conventions adopted for the northern hemisphere of the Earth, so that $p_z<0$ corresponds to the upward direction. b) 1D distribution function in the energy space (integrated over the pitch angle). The dotted line shows the distribution of the injected electrons (i.e., as if the wave-particle interaction was absent). The simulation parameters are given in Table \protect\ref{TableModel}.}
\label{FigEarth100}
\end{figure}
\begin{figure}
\includegraphics{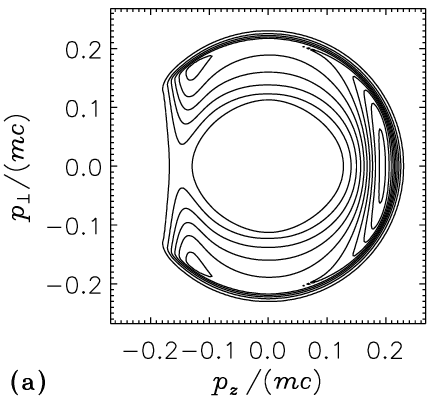}%
\includegraphics{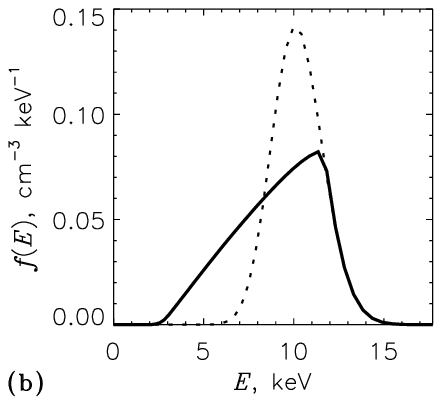}
\caption{Same as in Fig. \protect\ref{FigEarth100}, for the larger radiation source in the magnetosphere of the Earth (model \#2, $R_{\bot}=350$ km).}
\label{FigEarth350}
\end{figure}

\begin{figure}
\includegraphics{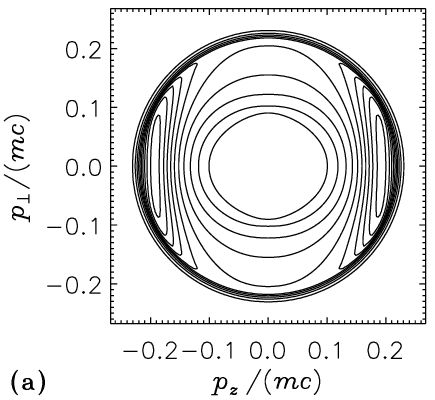}%
\includegraphics{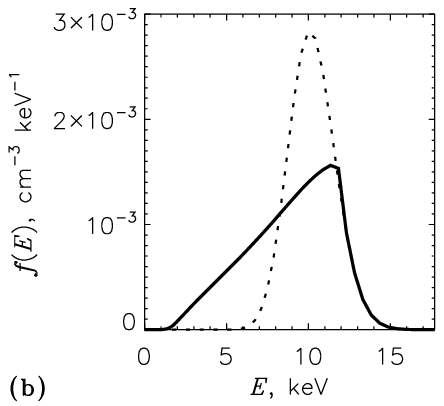}
\caption{Same as in Fig. \protect\ref{FigEarth100}, for the magnetosphere of Saturn.}
\label{FigSaturn}
\end{figure}

\begin{figure}
\includegraphics{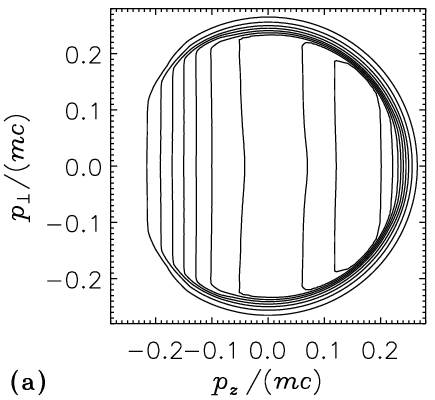}%
\includegraphics{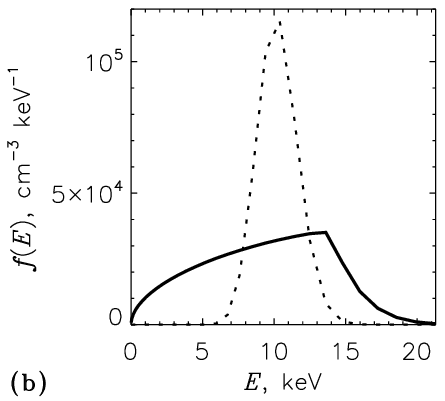}
\caption{Same as in Fig. \protect\ref{FigEarth100}, for the magnetosphere of an ultracool dwarf.}
\label{FigBD}
\end{figure}

Auroral kilometric radiation of the Earth is the best studied among the planetary radio emissions, with a lot of remote and in situ observations \citep[see, e.g., the reviews of][]{zar98, tre06}. The first model in Table \ref{TableModel} represents a typical source region with the transverse size of $\sim 100$ km, average emission intensity \citep{zar98}, and typical electron density and energy \citep{tre06}. The spectrum of the terrestrial radio emission covers a broad frequency range; we chose $\nu_{\mathrm{B}}=400$ kHz as a typical example. The simulated electron distribution in a quasi-stationary state\footnote{Note that the conversion from the 2D to the 1D distribution function shown in Figs. \ref{FigEarth100}-\ref{FigBD} includes multiplication by $p^2$.} is shown in Fig. \ref{FigEarth100}. One can see that the distribution is relatively weakly affected by the wave-particle interaction and does not differ significantly from the distribution of the injected electrons; the surface plot of the distribution function looks similar to that shown in Fig. \ref{FigBeam}b. The particles are concentrated in a relatively narrow energy range, which agrees with the observations \citep{tre06}. The conversion efficiency of the particle energy flux into waves amounts to only a few percent, but it is sufficient to provide the observed emission intensity (actually, the source size and the electron density and energy may be even smaller than in the considered example).

The second model of the source region of the terrestrial auroral radio emission in Table \ref{TableModel} corresponds to a particular event observed by the FAST satellite and reported by \citet{erg00}. This event is characterized by an unusually large source size ($R_{\bot}\sim 350$ km) and very high emission intensity (far exceeding the average values). We simulate this event using the same parameters as in the previous model, but with an increased transverse source size. As a result, the electron distribution in a quasi-stationary state is now more strongly relaxed (see Fig. \ref{FigEarth350}) and the conversion efficiency of the particle energy flux into waves reaches 11\%. Due to an increased maser efficiency (and larger source volume), the emission intensity is now considerably higher than in the previous model. We have found that by using the observed source dimensions and electron energy and density, we are able to reproduce the observed emission intensity. The calculated electron distribution also looks similar to those observed by FAST \citep{erg00}.

Recently, the Cassini spacecraft crossed the source region of Saturnian kilometric radio emission \citep{lam10, lam11, sch11}; the basic source parameters measured during this encounter are presented in Table \ref{TableObs}. These parameters, however, seem to be rather untypical because, firstly, they correspond to the low-frequency edge of the Saturnian radio emission spectrum \citep{zar98}. The observations were performed at a relatively large altitude; as a result, the loss-cone feature is almost absent ($\alpha_{\mathrm{c}}\sim 5^{\circ}$) and the particle escape timescale $\tau_{\mathrm{esc}}$ is much longer than that at the Earth. Secondly, the observed emission intensity far exceeds the average values \citep{zar98}. Nevertheless, by using the model based on the observed source dimensions and electron energy and density (see Table \ref{TableModel}), we are able to reproduce the observed emission intensity. The simulated quasi-stationary electron distribution (see Fig. \ref{FigSaturn}) agrees with the Cassini observations \citep{lam10, mut10}. This distribution is slightly more relaxed than that for the second model of the terrestrial auroral radio source; the conversion efficiency of the particle energy flux into waves reaches 13\%. We remind, however, that the considered case seems to be rather uncommon; in a more typical source of Saturnian auroral radio emission, the maser efficiency is expected to be much lower (similar to that at the Earth).

The nearby M9 dwarf TVLM 513 represents a ``classic'' example of a radio-emitting ultracool dwarf. To date, this object has been detected as a radio source in the $1.4-8.5$ GHz frequency range \citep{jae11}, although the actual spectrum may be broader. In addition to a quiescent component (which may be attributed to the gyrosynchrotron radiation), the radio emission includes short periodic pulses with high brightness temperature and $\sim 100\%$ circular polarization; the pulse period coincides with the stellar rotation period \citep{hal06, hal07, doy10, kuz11b}. Most likely, the periodic pulses are produced due to the electron-cyclotron maser instability, which requires a magnetic field of $\gtrsim 3000$ G. \citet{kuz11b} analyzed the radio light curves of TVLM 513 and identified the likely location of the emission source within the stellar magnetosphere (see Table \ref{TableObs}) by assuming a dipole magnetic field geometry. We assume that the transverse size of the emission source is similar to that at Saturn ($\sim 1000$ km), since the radii of the ultracool dwarf and the planet are comparable. We also assume that the typical electron energy in the magnetosphere of the ultracool dwarf is similar to that at the Earth and Saturn ($\sim 10$ keV), although it may actually be higher as suggested by the presence of a considerable gyrosynchrotron component in the radio emission. The electron injection rate $(\partial n_{\mathrm{e}}/\partial t)_{\mathrm{inj}}=5\times 10^6$ $\textrm{cm}^{-3}$ $\textrm{s}^{-1}$ was chosen so as to provide the observed emission intensity. The resulting electron density in a quasi-stationary state is $n_{\mathrm{e}}\simeq 4.2\times 10^5$ $\textrm{cm}^{-3}$. This is much higher than in the planetary magnetospheres, but the corresponding plasma-to-cyclotron frequency ratio ($\omega_{\mathrm{p}}/\omega_{\mathrm{B}}\simeq 10^{-3}$) is even lower than in the sources of auroral radio emissions of the Earth and Saturn, and is comparable with the estimations for Jupiter \citep{lec91, mel91}. The simulated quasi-stationary electron distribution (see Figs. \ref{FigBeam}f and \ref{FigBD}) is strongly relaxed and nearly flat. Therefore we expect the electron distributions in the magnetospheres of ultracool dwarfs to differ significantly from the shell- or horseshoe-like distributions that are typical of the source regions of planetary auroral radio emissions. Instead, we expect them to look similar to Maxwellian or kappa distributions, which are only slightly distorted by the parallel electric field and magnetic mirroring (i.e., the particle scattering on the waves is strong enough to keep the electron distribution close to an equilibrium state). Nevertheless, even these small deviations from an equilibrium distribution seem to be sufficient to produce an intense radio emission. 

The typical duration of radio pulses from TVLM 513 is about tens of seconds; moreover, the individual pulses seem to be caused by the star's rotation when a narrow radio beam sweeps periodically over an observer \citep{hal06, hal07, hal08, ber09, kuz11b}, while the parameters of the emission source region are stable at the timescales comparable to or exceeding the rotation period ($\simeq 2$ hours). These timescales far exceed the estimated timescale of the particle escape from the emission source ($\tau_{\mathrm{esc}}\sim 0.084$ s). Therefore the assumption that the electron distribution has reached a quasi-stationary state is well justified.

\section{Conclusion}\label{conclusion}
We have performed numerical simulations of the electron-cyc\-lo\-tron maser instability in a very low-density plasma. The used kinetic model included the relaxation of the electron distribution due to the wave-particle interaction, the injection of the energetic electrons with an unstable distribution into the emission source region, and the escape of the electrons from that region. A finite amplification time of the electromagnetic waves (caused by their escape from the source) was taken into account. The injected electrons were assumed to have the horseshoe-like distribution. We have found that:
\begin{itemize}
\item
The produced radio emission corresponds to the fundamental extraordinary mode, with the frequency slightly below the electron cyclotron frequency and the propagation direction nearly perpendicular to the magnetic field. This conclusion is valid both for the cases when the electron distribution is similar to that of the injected electrons, and when it is strongly relaxed. The intensity of other wave modes is negligible. In a relatively short time (which is determined by the particle escape time from the emission source region), the electron distribution and hence the emission intensity and spectrum reach a quasi-stationary state.
\item
Under the conditions typical of the sources of terrestrial and Saturnian auroral radio emissions, the dominant factor affecting the electron distribution is the particle escape from the emission source region. As a result, the electron distribution in a quasi-stationary state is weakly relaxed, i.e., it does not differ significantly from the horseshoe-like distribution of the injected electrons. The conversion efficiency of the particle energy flux into waves is typically a few percent, although it may be higher in particular events. The emission escape from the source region has a ``stabilizing'' effect on the electron distribution, i.e., it reduces the maser efficiency and allows the particles to have highly nonequilibrium but long-living (quasi-stationary) distributions. The simulated emission intensities and electron distributions agree well with the observations.
\item
Radio emission from ultracool dwarfs is much more intense than the planetary radio emissions. Therefore we expect that in the magnetospheres of ultracool dwarfs, the dominant factor affecting the electron distribution is the wave-particle interaction. As a result, the electron distribution in a quasi-stationary state is strongly relaxed and nearly flat; the conversion efficiency of the particle energy flux into waves is about 10\%. The radiation directivity pattern slightly differs from that for a shell- or horseshoe-driven instability; in particular, the emission is directed slightly downwards. The energetic electrons with a relatively low density ($\omega_{\mathrm{p}}/\omega_{\mathrm{B}}\sim 10^{-3}$) and energy of $\sim 10$ keV are able to provide the observed emission intensity.
\end{itemize}

\begin{acknowledgements}
A.A. Kuznetsov thanks the Leverhulme Trust for financial support. Research at Armagh Observatory is grant-aided by the Northern Ireland Department of Culture, Arts and Leisure.
\end{acknowledgements}

\bibliographystyle{aa}
\bibliography{ECMI}
\end{document}